\documentclass[]{elsart}
\usepackage{epsfig}

\usepackage{amsfonts}
\usepackage{amsbsy}
\usepackage{amssymb}

\def\be{\begin{equation}}
\def\ee{\end{equation}}
\def\ba{\begin{eqnarray}}
\def\ea{\end{eqnarray}}

\begin{document}

\begin{frontmatter}

\title{Multiscaling in Ising quantum chains with random 
Hilhorst--van Leeuwen perturbations}

\author{L. Turban}                    
 

\address{Laboratoire de Physique des Mat\'eriaux, UMR CNRS 7556,\\ 
Universit\'e  Henri Poincar\'e, Nancy 1,\\ 
F-54506 Vand\oe uvre l\`es Nancy Cedex, France}                                


\received{5 July 2001}


\begin{abstract}
We consider the influence on the surface critical behaviour of a 
quantum Ising chain of quenched random surface perturbations decaying 
as a power of the distance from the surface (random Hilhorst--van 
Leeuwen models). We study, analytically and numerically, the 
multiscaling behaviour of the surface magnetization and the 
surface energy density in the case of marginal perturbations. 
\end{abstract}


\end{frontmatter}

\section{Introduction}
\label{sec1}

In semi-infinite two-dimensional (2D) Ising models, extended perturbations 
of the coupling constants of the form $\Delta 
K_l=Al^{-\kappa}$, decaying as a power of the distance $l$ from 
a free surface, were first studied by Hilhorst and van Leeuwen 
(HvL) at the begining of the 1980s~\cite{hilhorst81}. Since then 
they have been the subject of continuous 
interest~\cite{cordery82,burkhardt82a,burkhardt82b,peschel84,bloete83,%
burkhardt84a,burkhardt84b,bloete85,bariev92,igloi93a,%
burkhardt90,berche90,turban93,karevski00}. Similar extended 
perturbations associated with line defects in the 
bulk~\cite{bariev88,bariev89,igloi90} or point 
defects~\cite{turban91,bariev91,peschel92} have also been considered 
(see Ref.~\cite{igloi93b} for a review). 

Such extended perturbations are known to modify the local critical 
behaviour, for an arbitrarily small value of the perturbation 
amplitude $A$, as soon as the decay exponent $\kappa$ is 
sufficiently small. Analytical results for a surface defect in the 
2D Ising model~\cite{hilhorst81,bloete83} show that 
with a temperature-like  perturbation (the case considered later) 
marginal behaviour is obtained when $\kappa=1$. 
Then the surface exponents vary continuously with the perturbation 
amplitude $A$ and the surface transition becomes first order when $A$ 
is greater than a critical value $A_{\rm c}$. When $\kappa<1$, the 
surface properties display essential singularities when $A<0$ and the 
transition is first order when $A>0$. 

This rich critical behaviour has  been explained independently by 
Cordery and Burkhardt~\cite{cordery82,burkhardt82a} through scaling 
considerations. Under a change of the length scale $l'=l/b$, the  
perturbation transforms as
\be 
[\Delta K_l]'=\frac{A'}{{l'}^\kappa}=b^{1/\nu}\frac{A}{l^\kappa}\;,
\label{e1.1}
\ee
where $\nu$ is the bulk correlation length exponent. Comparing both 
sides of the last equation, one obtains the following scaling 
behaviour for the perturbation amplitude:
\be 
A'=b^{-\kappa+1/\nu}A\,.
\label{e1.2}
\ee
Thus, for the 2D Ising model, $\kappa_{\rm c}=1/\nu=1$  corresponds to 
the critical value of the decay exponent below which the extended 
perturbation becomes relevant.

An interesting aspect of these smoothly inhomogeneous models is that 
in 2D their correlation functions are covariant under conformal 
transformations when the perturbation is marginal. Thus 
gap-exponent relations are 
satisfied~\cite{burkhardt90,turban93,igloi90,turban91,bariev91} and 
conformal profiles are obtained in the strip 
geometry~\cite{karevski00}, provided the form of the perturbation is 
also properly transformed as first noticed in Ref.~\cite{burkhardt90}.

Recently, aperiodic~\cite{turban99a} and 
random~\cite{turban99b,karevski99} versions of the HvL model have been
studied for the quantum Ising chain in a transverse field, which 
corresponds to the extreme anisotropic limit~\cite{kogut79} of the 2D 
classical Ising model and belongs to the  same universality class.

In Ref.~\cite{turban99b} random extended perturbations decaying towards a 
pure bulk system  were considered with two types of quenched 
randomness.

In model \#1, the random couplings take the following form:
\be
J_l=J\left[1+a_1\frac{(-1)^{f_l}}{l^\kappa}\right]\,,\qquad
f_l=\left\{\begin{array}{l}1\\ 0 \end{array}\right.
\quad{\rm with\ probability}\quad\begin{array}{l}1/2\\1/2\end{array}\,.
\label{e1.3}
\ee 
Thus the average perturbation $\langle{J_l-J}\rangle$ vanishes.

In model \#2
\be
J_l=J\left[1+f_la_2\right]\,,\qquad
f_l=\left\{\begin{array}{l}1\\ 0 \end{array}\right.
\quad{\rm with\ 
probability}\quad\begin{array}{l}l^{-\kappa}\\1-l^{-\kappa}\end{array}
\,.
\label{e1.4}
\ee
Then the average coupling $\langle{J_l}\rangle=J(1+a_2l^{-\kappa})$ has 
the same form as in the  HvL model.

In both models the transverse fields are kept constant, $h_l=h$.

A perturbation expansion of the average free energy and scaling 
considerations allowed us to obtain a relevance
criterion for both types of perturbations~\cite{turban99b}. With 
model \#1, the first-order correction to the average free energy 
vanishes and the second-order correction leeds to the following 
scaling behaviour for the amplitude $a_1$ of a thermal perturbation:
\be 
a_1'=b^{1/\nu-\kappa-1/2}a_1\,.
\label{e1.5}
\ee
With model \#2, the scaling behaviour of the perturbation amplitude 
$a_2$ is governed by the non-vanishing first-order correction to the 
average free energy so that
\be 
a_2'=b^{1/\nu-\kappa}a_2\,,
\label{e1.6}
\ee
as for the non-random HvL model in~(\ref{e1.2}).
According to~(\ref{e1.5}) and~(\ref{e1.6}) marginal behaviour is 
obtained for $\kappa\!=\!1/2$ with model \#1 and $\kappa\!=\!1$ with 
model \#2. 

Analytical expressions for the finite-size scaling behaviour at 
criticality of the average and typical surface magnetization were 
obtained in Ref.~\cite{turban99b}, in the case of marginal and relevant 
perturbations, in agreement with the results of numerical 
simulations. We gave also some conjectured expressions for
the surface energy exponents in the marginal case. 

In the present work, we extend our previous study by considering the 
finite-size scaling behaviour of the average of the moments of the 
surface magnetization $\langle{m_{\rm s}^q(L)}\rangle$ and of the 
singular part of the surface energy $\langle{e_{\rm s}^q(L)}\rangle$ 
in the marginal situation. In Section 2, we briefly recall the 
techniques used to diagonalize  the corresponding quantum Hamiltonians 
and to obtain the scaling behaviour of the surface magnetization and 
surface energy. Analytical expressions for the average of the moments 
of the surface magnetization and the surface energy are deduced from scaling 
considerations in Sections 3. The results are compared to numerical finite-size 
scaling data in Section 4 and their range of validity discussed in Section 5.

\section{Free fermions techniques}
\label{sec2}
\subsection{Diagonal Hamiltonian}
We consider the semi-infinite transverse-field Ising model (TIM) with 
Hamiltonian
\be 
{\mathcal H}=-\half\sum_{l=1}^\infty(J_l\sigma_l^z\sigma_{l+1}^z
+h_l\sigma_l^x)\,,
\label{e2.1}
\ee
where $\sigma_l^x$ and $\sigma_l^z$ are Pauli matrices, $J_l$ is the 
random coupling and $h_l$ the transverse field. 

The TIM Hamiltonian can be diagonalized using standard 
methods~\cite{lieb61,pfeuty70}. A Jordan--Wigner 
transformation~\cite{jordan28} changes the original 
Hamiltonian~(\ref{e2.1}) into a quadratic form in 
fermions which is diagonalized through a canonical transformation 
leading to
\be 
{\mathcal H}=\sum_{q=1}^L\epsilon_q
\left(\eta_q^\dagger\eta_q-\half\right)
\label{e2.2}
\ee
for a chain with length $L$. The $\eta_q^\dagger$ ($\eta_q$) are 
fermion creation (annihilation) operators and the non-negative 
excitation energies $\epsilon_q$ satisfy the set of equations
\ba 
\epsilon_q\psi_q(l)&=&-h_l\phi_q(l)-J_l\phi_q(l+1)\,,\nonumber\\
\epsilon_q\phi_q(l)&=&-J_{l-1}\psi_q(l-1)-h_l\psi_q(l)
\label{e2.3}
\ea 
with the boundary conditions $J_0=J_L=0$. Introducing $2L$-dimensional 
vectors $\boldsymbol{V}\!_q$ with components
\be 
V_q(2l-1)=-\phi_q(l)\,,\quad V_q(2l)=\psi_q(l)
\label{e2.4}
\ee 
the relations~(\ref{e2.3}) lead to an eigenvalue problem for the 
following tridiagonal matrix:
\be 
\mathsf{T}=\left( \begin{array}{cccccc}
\ \ 0\ \ &\ \ h_1\ \ & & & & \\
h_1 & 0 &\ \ J_1\ \ & & & \\
& J_1 & 0 &\  h_2\  & & \\
& & \ddots &\ \ \ddots\ \  & \ddots & \\
& & & J_{l-1} &\ \  0\ \  &  h_L \\
& & & & h_L &\ \ \ 0\ \   
\end{array} \right) \,.
\label{e2.5}
\ee 
When $\mathsf{T}$ is squared, odd and even components of 
$\boldsymbol{V}\!_q$ decouple and one obtains two separate eigenvalue 
problems for $\boldsymbol{\phi}$ and $\boldsymbol{\psi}$. The 
relations~(\ref{e2.3}) are invariant under the transformation 
$\boldsymbol{\phi}_q\rightarrow-\boldsymbol{\phi}_q$,
$\epsilon_q\rightarrow-\epsilon_q$ so that, changing 
$\boldsymbol{\phi}_q$ into $-\boldsymbol{\phi}_q$ in 
$\boldsymbol{V}\!_q$, one obtains the eigenvector of $\mathsf{T}$ 
with eigenvalue $-\epsilon_q$. Thus the physically relevant information 
is contained in that part of the spectrum of $\mathsf{T}$ with 
$\epsilon_q\geq0$.  

\subsection{Surface magnetization and surface energy}

The surface critical properties can be obtained through finite-size 
scaling at criticality, i.e., by working on a finite system with length 
$L$ and $J=h=1$. 

The imaginary time spin-spin autocorrelation function is given by
\be
{\mathcal G}_l^{\sigma\sigma}(\tau)=\langle0\vert\sigma_l^z(\tau)
\sigma_l^z(0)\vert0\rangle
=\sum_n\vert\langle 
n\vert\sigma_l^z\vert0\rangle\vert^2\exp[-\tau(E_n-E_0)]\,,
\label{e2.6}
\ee
where $\vert0\rangle$ and $\vert n\rangle$ are the ground state and the 
$n$th excited state of ${\mathcal H}$, $E_0$ and $E_n$ the 
corresponding eigenvalues. With fixed boundary conditions, 
$\sigma_L^z=\pm 1$, at $l=L$, the ground state is degenerate and 
asymptotically, one obtains  
$\lim_{\tau\to\infty}{\mathcal G}_l^{\sigma\sigma}(\tau)=m_l^2$ where
\be
m_l=\langle\sigma\vert\sigma_l^z\vert0\rangle
\label{e2.7}
\ee
is the local magnetization. It is given by an off-diagonal matrix 
element involving the first excited state with one fermionic 
excitation, $\vert\sigma\rangle=\eta^\dagger_1\vert0\rangle$, 
which is degenerate with the ground state. For the surface spin at 
$l=1$ a simple expression is obtained~\cite{peschel84}:
\be
m_{\rm s}(L)=m_1=\phi_1(1)=\left[1+
\sum_{l=1}^{L-1}\prod_{k=1}^l\left({h_k\over 
J_k}\right)^2\right]^{-1/2}\,.
\label{e2.8}
\ee
This expression can be rewritten under the following 
form~\cite{turban99b}:
\be
m_{\rm s}(L)=\overline{m}_{\rm s}^{\rm d}(L)
\prod_{l=1}^{L-1}\left({J_l\over h_l}\right)\,,
\label{e2.9}
\ee
where 
\be
\overline{m}_{\rm s}^{\rm d}(L)=\left[1+
\sum_{j=1}^{L-1}\prod_{k=L-j}^{L-1}\left({J_k\over
h_k}\right)^2\right]^{-1/2}
\label{e2.10}
\ee
is the surface magnetization at $l=L-1$ on the dual chain with 
transverse fields $J_l$, couplings $h_l$ $(l=0,L-1)$ and
fixed boundary conditions $J_0=0$ at $l=0$.

The scaling dimension of the surface energy density can be obtained by
considering the finite-size behaviour with free boundary conditions of 
the off-diagonal matrix element~\cite{berche96} 
\be 
e_{\rm s}(L)=\langle\varepsilon\vert\sigma_1^x\vert0\rangle
=(\epsilon_2-\epsilon_1)\phi_1(1)\phi_2(1)\,,
\label{e2.11}
\ee
where 
$\vert\varepsilon\rangle=\eta_1^\dagger\eta_2^\dagger\vert0\rangle$ is 
the lowest eigenstate leading to a non-vanishing matrix element. This 
matrix element enters into the expression of the connected 
energy-energy surface autocorrelation function 
${\mathcal G}_1^{\varepsilon\varepsilon}(\tau)$ and scales like the 
singular part of the surface energy density.

For the critical HvL model with a marginal decay of the perturbation, 
$h_l=1$, $J_l=1+a/l$, the product in~(\ref{e2.9}) behaves 
asymptotically as $L^a$. When $L\gg1$, the couplings on the right side 
of the chain are asymptotically unperturbed, thus one expects that 
$\overline{m}_{\rm s}^{\rm d}(L)\sim L^{-1/2}$ as for the homogeneous 
Ising model at the ordinary transition. This leads to the scaling 
behaviour:
\be
 m_{\rm s}(L)\sim L^{-x_{\rm m}^{\rm s}}\,,\qquad
x_{\rm m}^{\rm s}=\half-a\,,\qquad a\leq\half\,.
\label{e2.12}
\ee
This expression cannot be valid beyond $a=1/2$ where the surface 
remains ordered at the bulk critical point~\cite{bloete83,peschel84}. 
The surface magnetization then displays a first order transition and 
$x_{\rm m}^{\rm s}=0$ when $a>a_c=1/2$. Thus our assumption concerning 
the behaviour of 
$\overline{m}_{\rm s}^{\rm d}(L)$ is wrong when $a>1/2$ and then one 
must have $\overline{m}_{\rm s}^{\rm d}(L)\sim L^{-a}$. 

For the surface energy in Eq.~(\ref{e2.11}), when $a\leq1/2$ both
$\phi_1(1)$ and $\phi_2(1)$ scale as $m_{\rm s}(L)\sim L^{-x_{\rm 
m}^{\rm s}}$. The excitations $\epsilon_1$ and $\epsilon_2$ scale as 
$L^{-1}$ leading to 
\be
e_{\rm s}(L)\sim L^{-1}m_{\rm s}^2(L)\sim L^{-x_{\rm e}^{\rm s}} 
\,,\qquad x_{\rm e}^{\rm s}=2(1-a)\,,\qquad a\leq\half\,,
\label{e2.13}
\ee
in the second-order regime. When $a>1/2$, the onset of surface order 
leads to a more complicated scaling behaviour. The first gap 
$\epsilon_1$ is associated with a localized state which reflects the 
surface behaviour. It scales as $L^{-2a}$ while $\epsilon_2\sim L^{-1}$ 
as usual. We have $\phi_1(1)=m_{\rm s}(L)\sim L^0$ but $\phi_2(1)$ 
behaves anomalously as $L^{-a+1/2}$. Finally, $x_{\rm e}^{\rm s}=a+1/2$ 
in the first order regime, $a>1/2$.

\section{Calculation of the moments}
\label{sec3}

In this section we evaluate the moments of the surface magnetization 
and the surface energy density at criticality on a chain with length 
$L$ in the regime of second-order surface transition. We restrict our 
study to the case of marginal perturbations, i.e., with $\kappa=1/2$ 
for model \#1 and $\kappa=1$ for model \#2. 

\subsection{Surface magnetization}

We assume that the surface magnetization at $l=L-1$ on the
dual chain, $\overline{m}_{\rm s}^{\rm d}(L)$ in~(\ref{e2.9}), is 
still scaling as $L^{-1/2}$ for the random models so that
\be
\langle m_{\rm 
s}^q(L)\rangle=L^{-q/2}\langle\prod_{l=1}^{L-1}J_l^q\rangle
=L^{-q/2}\prod_{l=1}^{L-1}\langle J_l^q\rangle\,,
\label{e3.1}
\ee
where we used the statistical independence of the couplings on 
different bonds. Using the distribution of the
couplings given in~(\ref{e1.3}) for model~\#1, with $J=1$ we obtain
\be
\langle J_l^q\rangle=\half\left(1+{a_1\over l^{1/2}}\right)^q
+\half\left(1-{a_1\over l^{1/2}}\right)^q
\simeq 1+{q(q-1)\over 2}{a_1^2\over l}\,.
\label{e3.2}
\ee
The last expression gives the asymptotic behaviour when $l\gg 1$ which 
governs the critical behaviour. Taking the logarithm of the product 
in~(\ref{e3.1}) and a continuum approximation leads to
\be 
\ln\prod_{l=1}^{L-1}\langle J_l^q\rangle
\simeq \int_1^L\ln\left(1+{q(q-1)\over 2}{a_1^2\over l}\right)\d l
\simeq {q(q-1)a_1^2\over 2}\ln L\,.
\label{e3.3}
\ee
Thus the $q$th moment of the surface magnetization scale as
\be
\langle m_{\rm s}^q(L)\rangle\sim L^{-qx_{\rm m}^{\rm s}(q)}
\sim L^{-(q/2)[1-(q-1)a_1^2]}
\label{e3.4}
\ee
and the surface magnetization displays multiscaling with
\be
x_{\rm m}^{\rm s}(q)=\half[1-(q-1)a_1^2]\qquad
(\rm model\ \#1)\,.
\label{e3.5}
\ee

Similarly for model~\#2, using~(\ref{e1.4}) with $J=1$ and $\kappa=1$, 
we obtain
\be
\langle J_l^q\rangle={1\over l}(1+a_2)^q+1-{1\over l}
=1+{(1+a_2)^q-1\over l}\,.
\label{e3.6}
\ee
Since
\be
\ln\prod_{l=1}^{L-1}\langle J_l^q\rangle
\simeq \int_1^L\ln\left[1+{(1+a_2)^q-1\over l}\right]\d l
\simeq [(1+a_2)^q-1]\ln L\,,
\label{e3.7}
\ee
the $q$th moment of the surface magnetization scales as
\be
\langle m_{\rm s}^q(L)\rangle\sim L^{-(q/2)+(1+a_2)^q-1}\,,
\label{e3.8}
\ee
so that
\be
x_{\rm m}^{\rm s}(q)={1\over 2}-{(1+a_2)^q-1\over q}\qquad
(\rm model\ \#2)\,.
\label{e3.9}
\ee

The exponents in Eqs.~(\ref{e3.5}) and~(\ref{e3.9}) are in 
agreement with the average and typical exponents obtained previously 
in Ref.~\cite{turban99b} when $q=1$ and $q\to 0$, respectively:
\be
\begin{array}{lll}
{[x_{\rm m}^{\rm s}]}_{\rm av}=\half\,,\qquad
&{[x_{\rm m}^{\rm s}]}_{\rm typ}=\half(1+a_1^2)\qquad &(\rm model\ 
\#1)\\
{[x_{\rm m}^{\rm s}]}_{\rm av}=\half-a_2\,,\qquad
&{[x_{\rm m}^{\rm s}]}_{\rm typ}=\half-\ln(1+a_2)\qquad &(\rm model\ 
\#2)
\end{array}
\label{e3.10}
\ee

\subsection{Surface energy density}

For the scaling behaviour of the moments of the surface energy density, 
one can start from the expression giving $e_{\rm s}(L)$ in 
Eq.~(\ref{e2.13}). The factor $L^{-1}$ is associated with the 
scaling of the excitations and reflects the isotropy of the bulk 
critical behaviour. It should not be modified with a random surface 
perturbation, hence we have
\be
\langle e_{\rm s}^q(L)\rangle\sim L^{-q}
\langle m_{\rm s}^{2q}(L)\rangle\,.
\label{e3.11}
\ee
The $q$th moment of the surface energy density depends on the moment of 
order $2q$ of the surface magnetization. 

For model \#1, making use of Eq.~(\ref{e3.4}), we obtain
\be
\langle e_{\rm s}^q(L)\rangle\sim L^{-qx_{\rm e}^{\rm s}(q)}
\sim L^{-q[2-(2q-1)a_1^2]}\,,
\label{e3.12}
\ee
so that 
\be
x_{\rm e}^{\rm s}(q)=2-(2q-1)a_1^2\qquad
(\rm model\ \#1)\,.
\label{e3.13}
\ee

In the same way Eq.~(\ref{e3.8}) leads to
\be
\langle e_{\rm s}^q(L)\rangle\sim L^{-2q+(1+a_2)^{2q}-1}
\label{e3.14}
\ee
for model \#2 so that
\be
x_{\rm e}^{\rm s}(q)=2-{(1+a_2)^{2q}-1\over q}\qquad
(\rm model\ \#2)\,.
\label{e3.15}
\ee

The average and typical exponents are then given by:
\be
\begin{array}{lll}
{[x_{\rm e}^{\rm s}]}_{\rm av}=2-a_1^2\,,\quad
&{[x_{\rm e}^{\rm s}]}_{\rm typ}=2+a_1^2\quad &(\rm model\ \#1)\\
{[x_{\rm e}^{\rm s}]}_{\rm av}=2-a_2(2+a_2)\,,\quad
&{[x_{\rm e}^{\rm s}]}_{\rm typ}=2[1-\ln(1+a_2)]\quad &(\rm model\ \#2)
\end{array}
\label{e3.16}
\ee
The typical exponents in Eqs.~(\ref{e3.10}) and~(\ref{e3.16}) are 
in agreement with the values conjectured in Ref.~\cite{turban99b}. 
They keep the same form as for the pure HvL model in 
Eqs.~(\ref{e2.12}) and~(\ref{e2.13}) with $a$ replaced by an 
effective amplitude, $-a_1^2/2$ for model \#1 and  $\ln(1+a_2)$ for 
model \#2. These typical effective amplitudes are deduced from the 
relation $\ln J_l(a_{\rm eff})\simeq \langle\ln J_l\rangle$ for 
$l\gg1$. A similar conjecture proposed in Ref.~\cite{turban99b} for the 
average exponents is actually verified only up to the first order in 
the perturbation amplitude.

\section{Numerical results}
\label{sec4}

\subsection{Surface magnetization}

The surface magnetization for a given configuration of the couplings 
$\{J_k\}$ was obtained using Eq.~(\ref{e2.8}) at criticality, 
i.e., with $h_k=1$ and $J=1$ in~(\ref{e1.3}) and~(\ref{e1.4}). 

\begin{figure}[thb]
\centerline{\psfig{figure=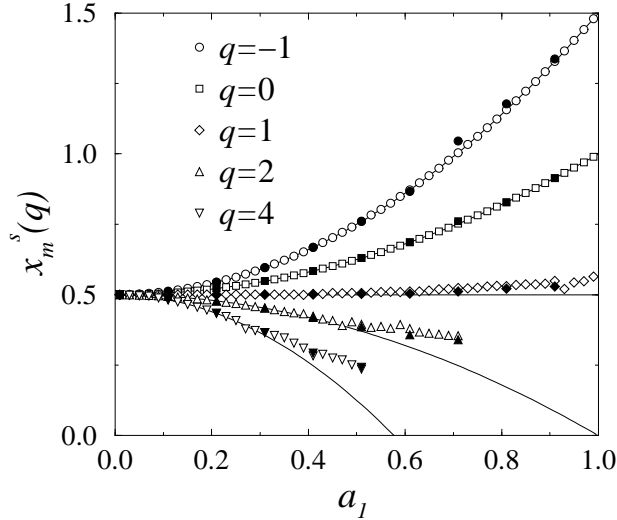,width=8.0cm,angle=0}}
\vglue-.3cm
\caption{Exponent of the $q$th moment of the surface magnetization as a 
function of the perturbation amplitude for model \#1. The  extrapolated 
exponents were deduced either from exact enumerations for chains with 
sizes $L=4$--$24$ (open symbols) or from Monte Carlo samplings for 
chains with sizes $L=2^5$--$2^{14}$ (filled symbols). The lines 
correspond to the analytical expression in (\ref{e3.5}).} 
\label{fig1}
\end{figure}
\medskip

\begin{figure}[hbt]
\centerline{\psfig{figure=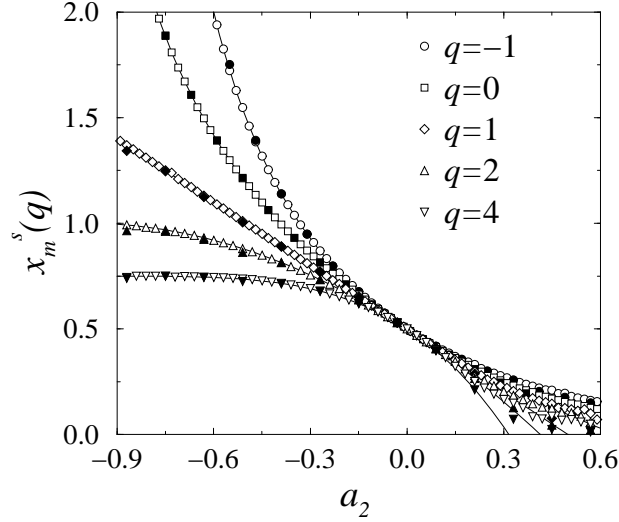,width=8.0cm,angle=0}}
\vglue-.3cm
\caption{As in Fig.~\ref{fig1} for model \#2. The lines correspond to 
the analytical expression in (\ref{e3.9}).} 
\label{fig2}
\end{figure}
\medskip

The moments were averaged either by performing exact enumerations of 
the random configurations in the case of chains with lengths 
$L=4$--$24$ or through Monte Carlo samplings (over $10^6$ samples) for 
longer chains with lengths $L=2^5$--$2^{14}$. The exponents
$x_{\rm m}^{\rm s}(q)$ were deduced from an extrapolation of two-point 
approximants using the BST algorithm~\cite{henkel88}.   
The marginal exponents are shown in Fig.~\ref{fig1} for model~\#1 and 
in Fig.~\ref{fig2} for model~\#2.  
 
\subsection{Surface energy density}

\begin{figure}[bht]
\centerline{\psfig{figure=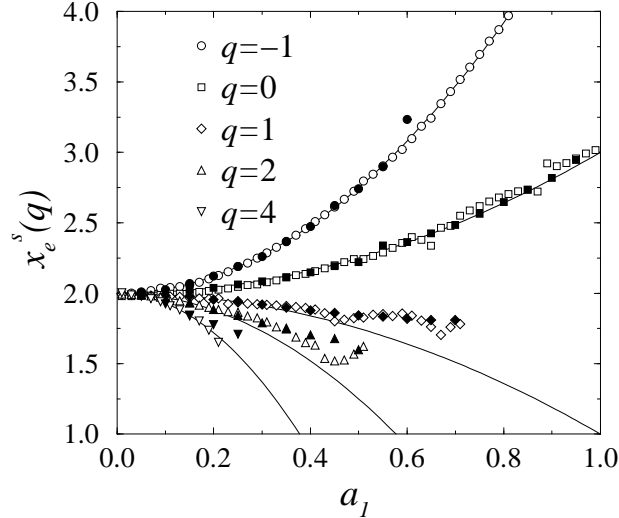,width=8.0cm,angle=0}}
\vglue-.3cm
\caption{Exponent of the $q$th moment of the surface energy density as 
a function of the perturbation amplitude for model \#1. The  
extrapolated exponents were deduced either from exact enumerations for 
chains with sizes $L=4$--$24$ (open symbols) or from Monte Carlo 
samplings for chains with sizes $L=2^2$--$2^{9}$ (filled symbols). The 
lines correspond to the analytical expression in (\ref{e3.13}).} 
\label{fig3}
\end{figure}
\medskip

\begin{figure}[thb]
\centerline{\psfig{figure=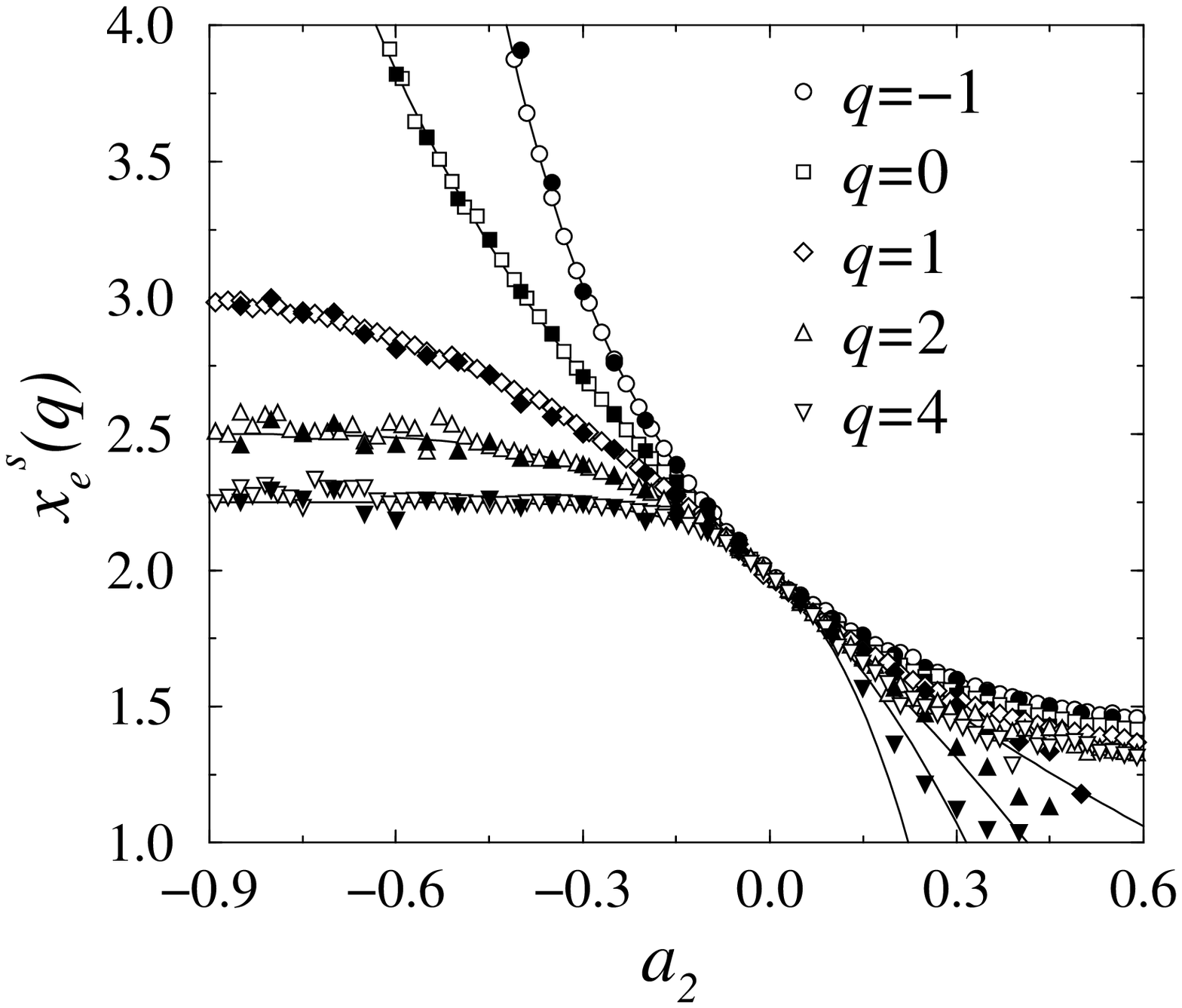,width=8.0cm,angle=0}}
\vglue-.3cm
\caption{As in Fig.~\ref{fig3} for model \#2. The lines correspond to 
the analytical expression in (\ref{e3.15}).} 
\label{fig4}
\end{figure}
\medskip

The surface energy density was obtained using the expression given 
in~(\ref{e2.11}) where $\epsilon_1$, $\epsilon_2$, $\phi_1(1)$ and 
$\phi_2(1)$ are obtained through the diagonalisation of a $L\times L$ 
matrix constructed from the elements of $\mathsf{T}^2$ with odd 
indices. 

The average moments were deduced from exact enumerations for chains 
with lengths $L=4$--$24$ and from Monte Carlo samplings (over $10^4$ 
samples) for chains with lengths $L=2^2$--$2^{9}$. The exponents 
$x_{\rm e}^{\rm s}(q)$ were deduced from an extrapolation of two-point 
approximants using the BST algorithm in the case of exact enumerations. 
The approximants deduced from the Monte Carlo data were too noisy to 
use the BST extrapolation process. In this case the exponents were 
deduced from a non-linear fit of the following expression:
\be
\ln\langle e_{\rm s}^q(L)\rangle\simeq 
\ln A-x_{\rm e}^{\rm s}(q) \ln L+BL^{-\omega}\,.
\label{e4.1}
\ee
The continuously varying exponents are compared to the analytical 
expressions in Fig.~\ref{fig3} for model~\#1 and in Fig.~\ref{fig4} 
for model~\#2.  

\section{Discussion}
\label{sec5}

\begin{figure}[bht]
\centerline{\psfig{figure=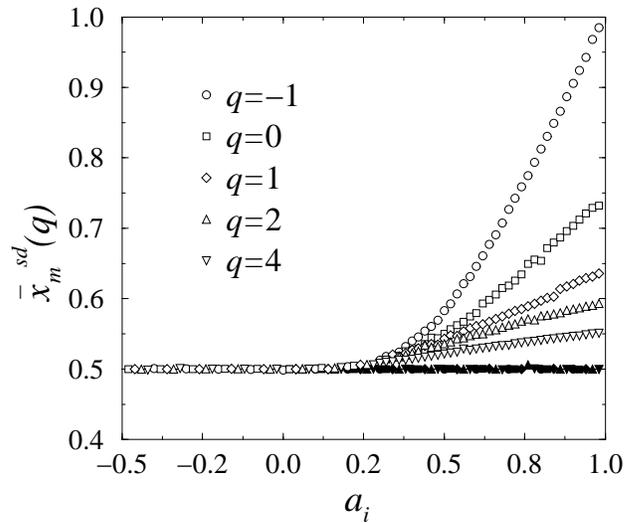,width=8.0cm,angle=0}}
\vglue-.3cm
\caption{Exponent of the $q$th moment of the dual surface magnetization 
at $L$ as a function of the perturbation amplitude for model \#1 (full 
symbols) and model \#2 (open symbols). The extrapolated exponents were 
deduced from exact enumerations for chains with sizes $L=4$--$24$. The 
line corresponds to the behaviour at the ordinary surface transition 
where $x_{\rm m}^{\rm s}=1/2$.} 
\label{fig5}
\end{figure}
\medskip

A good agreement between numerical and conjectured analytical results have 
been obtained for the surface magnetization and the surface energy 
density, except in the vicinity of first-order surface transitions. 

Already for the pure HvL model a slow convergence to the exactly known 
behaviour was observed close to the first-order regime~\cite{berche90}. 
One may verify on Figs.~\ref{fig2} and~\ref{fig4} for model \#2 
that the Monte Carlo data, obtained on larger systems, are closer to the 
analytical results when $a_2>0$. 
Nevertheless, like for the the pure system, some of the scaling 
assumptions leading to the analytical expressions for the surface 
exponents cannot be valid beyond some critical value of the perturbation 
amplitude since $x_{\rm m}^{\rm s}(q)$ and $x_{\rm e}^{\rm s}(q)$ 
have to remain non-negative. 

The dual magnetization in~(\ref{e2.9}) was assumed to scale as $L^{-1/2}$, 
like for the unperturbed system at the ordinary transition. We performed a 
numerical study of the moments of $\overline{m}_{\rm s}^{\rm d}(L)$ for 
the two models using Eq.~(\ref{e2.10}), exact enumerations for chains 
with lengths $L=4$--$24$ and BST extrapolations. The results are shown in 
Fig.~\ref{fig5}. 

For model \#1, the ordinary surface behaviour is obtained for any value of 
the perturbation amplitude which gives some confidence in the conjectured 
analytical expressions in the regime of second-order transition. 
For model \#2, the ordinary surface behaviour is only obtained when 
$a_2\lesssim.2$. Beyond this value deviations from the ordinary surface 
behaviour are observed and the analytical expressions for the exponents are 
probably no longer valid.

\ack
Useful discussions with Dragi Karevski and Ferenc Igl\'oi are 
gratefully acknowledged.

\end{document}